\documentclass[reprint,
superscriptaddress,
 amsmath,amssymb,
 aps,prl
]{revtex4-2}
\usepackage{graphicx}
\usepackage{dcolumn}
\usepackage{bm}
\usepackage{xcolor}

\begin{document}

\preprint{APS/123-QED}

\title{Optical trapping of high-aspect-ratio NaYF hexagonal prisms for kHz-MHz gravitational wave detectors}

\author{George Winstone}
\altaffiliation{These authors contributed equally}
\affiliation{Center for Fundamental Physics, Department of Physics and Astronomy,Northwestern University, Evanston, Illinois 60208, USA}

\author{Zhiyuan Wang}
\altaffiliation{These authors contributed equally}
\affiliation{Center for Fundamental Physics, Department of Physics and Astronomy,Northwestern University, Evanston, Illinois 60208, USA}

\author{Shelby Klomp}%
\altaffiliation{These authors contributed equally}
\affiliation{Center for Fundamental Physics, Department of Physics and Astronomy,Northwestern University, Evanston, Illinois 60208, USA}

\author{Greg Felsted}%
\altaffiliation{These authors contributed equally}
\affiliation{ Department of Chemistry, University of Washington, Seattle, Washington 98195, USA}

\author{Andrew Laeuger}
\affiliation{Center for Fundamental Physics, Department of Physics and Astronomy,Northwestern University, Evanston, Illinois 60208, USA}

\author{Daniel Grass}%
\affiliation{Center for Fundamental Physics, Department of Physics and Astronomy,Northwestern University, Evanston, Illinois 60208, USA}

\author{Nancy Aggarwal}%
\affiliation{Center for Fundamental Physics, Department of Physics and Astronomy,Northwestern University, Evanston, Illinois 60208, USA}
\affiliation{Center for Interdisciplinary Exploration and Research in Astrophysics (CIERA), Department of Physics and Astronomy, Northwestern University, Evanston, Illinois 60208, USA}

\author{Jacob Sprague}%
\affiliation{Center for Interdisciplinary Exploration and Research in Astrophysics (CIERA), Department of Physics and Astronomy, Northwestern University, Evanston, Illinois 60208, USA}

\author{Peter J. Pauzauskie}%
\affiliation{Department of Materials Science, University of Washington, Seattle, Washington 98195, USA}
\affiliation{Physical Sciences Division, Physical and Computational Sciences Directorate, Pacific Northwest National Laboratory, Richland, Washington 99352, USA}

\author{Shane L. Larson}%
\affiliation{Center for Interdisciplinary Exploration and Research in Astrophysics (CIERA), Department of Physics and Astronomy, Northwestern University, Evanston, Illinois 60208, USA}

\author{Vicky Kalogera}%
\affiliation{Center for Interdisciplinary Exploration and Research in Astrophysics (CIERA), Department of Physics and Astronomy, Northwestern University, Evanston, Illinois 60208, USA}

\author{Andrew A. Geraci}%
\affiliation{Center for Fundamental Physics, Department of Physics and Astronomy,Northwestern University, Evanston, Illinois 60208, USA}
\affiliation{Center for Interdisciplinary Exploration and Research in Astrophysics (CIERA), Department of Physics and Astronomy, Northwestern University, Evanston, Illinois 60208, USA}

\collaboration{And the LSD Collaboration}

\date{\today}

\begin{abstract}
We present experimental results on optical trapping of Yb-doped $\beta-$NaYF sub-wavelength-thickness high-aspect-ratio hexagonal prisms with a micron-scale radius. The prisms are trapped in vacuum using an optical standing wave, with the normal vector to their face oriented along the beam propagation direction. 
The measured motional spectra are compared with numerical simulations. 
This plate-like geometry simultaneously enables trapping with low photon-recoil-heating, high mass, and high trap frequency, potentially leading to advances in high frequency gravitational wave searches in the Levitated Sensor Detector (LSD), currently under construction \cite{aggarwal2022searching}.
The material used here has previously been shown to exhibit internal cooling via laser refrigeration when optically trapped and illuminated with light of suitable wavelength \cite{Nick2021,Zhou2016}. Employing such laser refrigeration methods in the context of our work may enable higher trapping intensity thus and higher trap frequencies for gravitational wave searches approaching the several hundred kHz range. 
\end{abstract}

\maketitle




The field of levitated optomechanics is both rapidly developing and of high scientific interest, with a number of impressive recent results including achieving cooling to the quantum ground state \cite{Uros,Tebbenjohanns2021}, high resolution surface force mapping \cite{blakemore2019three,montoya2021scanning,winstone2018direct}, material limited GHz rotations \cite{reimann2018ghz}, microscopic material studies \cite{ricci2021chemical}, and high precision force \cite{ranjit2016zeptonewton} and acceleration \cite{monteiro2017optical} sensitivity.
Near term goals include contributions to dark matter and energy searches \cite{monteiro2020search} and astrophysics, including searches for high frequency gravitational waves (GWs) in the Leviated Sensor Detector (LSD), currently under construction \cite{aggarwal2022searching}.

With the first detection of gravitational waves \cite{LIGOfirst2016}, there have been a number of proposed and constructed experimental efforts to extend the observable GW frequency spectrum, such as the upcoming LISA mission \cite{LISA1,LISA2} 
 and various other proposals \cite{GWclocks,MAGIS,holometer,Akutsu:2008qv,domcke2021potential,Singh:2016xwa,cruise2012potential,vermeulen2021experiment,nishizawa2008laser,ito2020probing,ejlli2019upper}.  Levitated sensors have been identified as a promising route to search for gravitational waves in the range of 10 kHz to a few hundred kHz \cite{Arvanitaki:2013,aggarwal2022searching}, where sources can include GW emission from annihilation of axions in clouds around spinning black holes 
 or from inspirals and mergers of nearby sub-solar-mass primordial black holes within the Milky Way. 
The sensitivity of such a levitated sensor detector benefits from low photon-recoil-heating and  greater mass of the levitated particle. Both can be enabled by trapping disc-like objects instead of spheres \cite{aggarwal2022searching}.

In this Letter, we demonstrate optical trapping and study the motional dynamics of high aspect ratio, high mass, high mechanical frequency hexagonal prisms in vacuum. We compare the measured motional spectra with simulations, and we illustrate the potential advances for high frequency gravitational wave detection made possible by trapping particles of this geometry.
The main advance here is in achieving optical trapping and characterizing the dynamics of a dielectric object with a low-photon-recoil geometry while maintaining a large mass and high trapping frequency, which is essential for realizing the design sensitivity for the LSD   gravitational wave search at frequencies above 10 kHz \cite{aggarwal2022searching}. 

While small spherical particles can be trapped at high frequency ($\sim 300$ kHz) \cite{Uros,Tebbenjohanns2021}, the near-isotropic nature of their light scattering yields significant photon recoil heating, and their low mass is undesirable for the application of GW detection. Larger diameter ($\sim 10$ $\mu$m) spherical particles have significant mass, but have only been realized in sub-kHz-frequency trapping configurations \cite{monteiro2017optical,Rider2016}. The disc-like or plate-like geometry of the high-aspect-ratio hexagonal prisms we study in this work allows the ideal combination of high mass, high frequency, and low photon-recoil-heating, and exhibits a clear improvement over levitated spheres, both in the
regime when the sensor is dominated by thermal noise
and when the sensor is limited by photon recoil.
While these objects are of a lower mass than in our ideal design of LSD \cite{aggarwal2022searching}, this work represents a significant step along the technical roadmap and prepares us for trapping more customized similar objects in the final detector.

%
%
%
%
Since the first experimental demonstration of cold Brownian motion \cite{Roder2015}, solid-state laser refrigeration \cite{Xia2021} has proven to be an effective way of preventing detrimental photothermal heating of optically-levitated materials. In addition, NaYF and other rare-earth-doped crystals have been studied before in the context of optical vacuum levitation \cite{Rahman2017,Nick2021} and have been shown to exhibit laser refrigeration when trapped with light of an appropriate wavelength. 
Previous experiments with both cubic ($\alpha$) and hexagonal ($\beta$) phases of NaYF have either been with low mass subwavelength particles \cite{Nick2021} or not optically levitated in vacuum \cite{Zhou2016}.
By choosing such a material for the sensor, in the future laser refrigeration methods may enable higher trapping intensity and higher trap frequencies for gravitational wave searches approaching the several hundred kHz frequency range.

Finally, as a general tool for precision sensing experiments, high aspect ratio trapped objects enable levitation of a multi-micron scale object with mechanical frequencies in the tens of kHz, much higher than what is possible for similar mass spherical objects at similar optical powers \cite{monteiro2017optical,Rider2016}, due to the sub-wavelength size of the object in the optical axis. 
Additionally, $\beta$-NaYF has a density roughly double that of $\text{SiO}_2$, further increasing the mass of the trapped particle.
Due to the presence of common noise sources at lower frequencies for example from ground vibration, seismic activity, or acoustic noise, this is a useful experimental platform for precision measurements, e.g. involving accelerometry, that would benefit from a sensor with larger mass and higher bandwidth.


\begin{figure}[h]
\includegraphics[width=0.98\columnwidth]{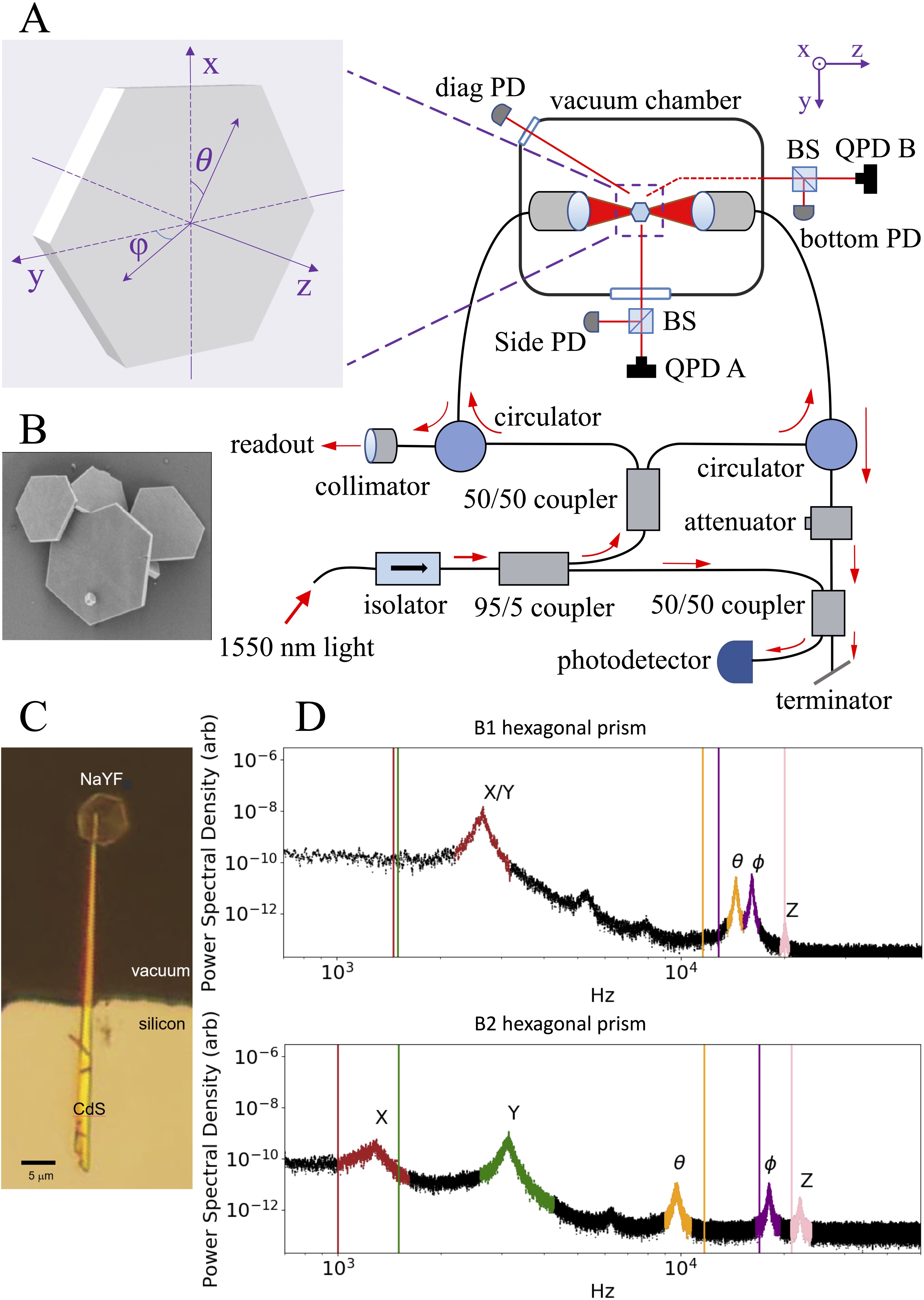}  
\caption{\label{fig:optics_layout} (a) Setup. The hexagonal prism is trapped at the conjoined foci of two 1550 nm counter-propagating linearly polarized beams. Fiber-based detection is achieved through homodyne-like interference of light coupled back into the PM fiber and the reference light out of the 95/5 coupler through a fiber photodetector (bandwidth of 1.6 GHZ).  Scattered light is also  detected through a series of three biased photodiodes (PDs) and two quadrant photodetectors (QPDs) viewing the prism from the side, bottom, and top diagonal directions, as indicated. (b) Scanning electron beam micrograph of a cluster of NaYF hexagonal prisms from the $B2$ sample. (c) Optical micrograph of NaYF prism, adapted from Ref. \cite{petercrystalgrowth}. (d) Spectral density of the motional signal at $2.2$ mbar recorded by one of the QPD detectors for two different size hexagonal prisms, batch $B1$ (top $P=475$ mW) and $B2$ (bottom $P=250$ mW), respectively, with dimensions as described in the text. Motion corresponding to five principal degrees of freedom $(x,y,z,\theta, \phi)$ of the hexagon is visible, due to the non-perfectly-orthogonal viewing angle of the detector. Vertical lines indicate calculated frequencies from a finite-element simulation for similar prisms, showing qualitative agreement.}
\end{figure}

\textit{Experimental Setup --}
A top-view schematic of the experimental setup is shown in Fig. \ref{fig:optics_layout}a. Two focused counter-propagating linearly polarized Gaussian beams, which are obtained from a 1550 nm laser via polarization-maintaining (PM) fiber-optic couplers and circulators, create a standing-wave optical dipole trap. Yb-doped $\beta$-NaYF hexagonal prisms are released from a glass substrate when driven by a piezoelectric transducer under 2 - 12 mbar of $N_2$ gas \cite{Beckthesis} and trapped in one of the anti-nodes of the standing wave (total power at the trap $P$ $\approx$ $475$ mW, and beam waist $w$ $\approx$ 12$\mu$m). We study two batches of hexagonal prisms which we label as B1 and B2, where the aspect ratios are approximately 8:1 (2.5 $\mu$m in diameter and 300 nm thick) and 15 - 25:1 (3 - 5 $\mu$m in diameter and 200 nm thick), respectively. The $\beta$-NaYF crystals are hydrothermally grown in an autoclave, resulting in a distribution of particle radii and thicknesses (see Supplementary Material).

To characterize the motional dynamics of the prisms in the trap, 
we employed multiple independent detection mechanisms, as illustrated in Fig. \ref{fig:optics_layout}a. Light scattered from the prism is partially coupled back into the PM fiber and interferes with the 
reference light out of the 95/5 optical coupler to achieve homodyne-like interferometric detection through a fiber photo-detector. The image of the prism is projected onto two quadrant photo-detectors (QPDs) from the side and bottom of the vacuum chamber respectively. Additionally, scattered light is directly detected by three free-space biased photodiodes (PDs) which view the prism from the side, bottom, and top diagonal windows of the chamber. 
Along the optical axis the positional information of the prism is encoded mainly in the phase of the light given the prism's sub-wavelength thickness in this direction.  Off-axis positional information is encoded mainly in the intensity due to the much larger (few $\mu$m) transverse dimensions of the prism.

\textit{Theoretical Model--} Optical trapping of spheres in the Rayleigh, Mie-Lorentz and geometric optics regime is a fairly well studied problem experimentally and theoretically. Furthermore, several novel geometries have been explored in the sub-wavelength (Rayleigh) regime.
The expected form for non spherical objects optically trapped in the Mie-Lorentz regime is a topic of recent theoretical investigation \cite{seberson2020stability}.
While a sphere trapped in the Mie-Lorentz regime will exhibit 3 degrees of freedom (DOF), in contrast for a high-aspect-ratio radially symmetrically object (i.e a disk or disk like object) we would expect to see 5 degrees of freedom, since the 6th DOF (that of the hexagonal prism spinning around its most symmetrical axis) is poorly optically coupled. 
 Our $\beta-$NaYF hexagons are neither perfectly symmetric nor optically isotropic so we would expect to see a coupling to the output light in the case that the rotations around that axis were driven. This could be explored in future experiments for example by using circularly polarized trapping light, since $\beta$-NaYF is birefringent. 

Approximate, first-order solutions for a thin disc are given in \cite{seberson2020stability}. The analytic approach predicts the mode ordering in frequency space that we observe, 
however due to the non-infinitesimal thickness of the hexagonal prisms used in the experiment we use a finite element model implemented in PYGDM2 \cite{pyGDM2} to compute the expected optical forces and frequencies of the trapped prisms. Computed frequencies for our geometries are overlaid over experimental data in Fig. \ref{fig:optics_layout}d.





\begin{figure}[h]
\includegraphics[width=1.0\columnwidth]{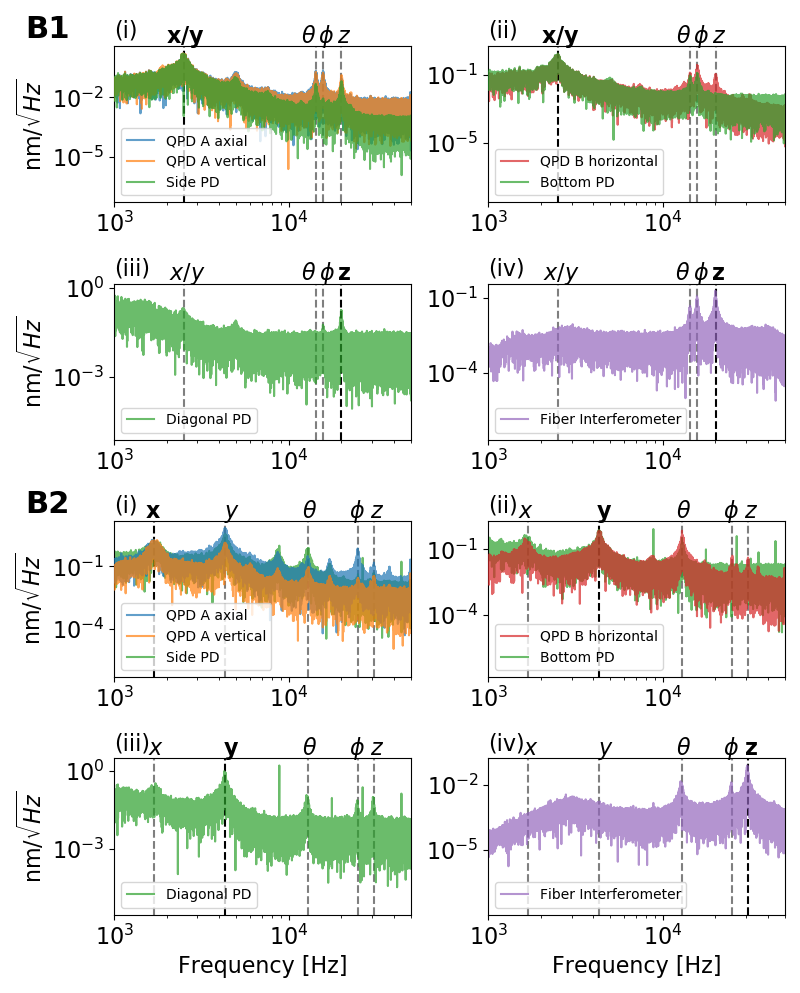}
\caption{\label{fig:all_data}Experimentally obtained power spectral densities for two sizes of hexagonal prisms, B1 (top) and B2 (bottom) from multiple detectors oriented to the side (i), bottom (ii), top diagonal (iii), and along the trap axis (iv). Detector types include quadrant photo-detectors (blue, orange, and red), biased photodiodes (green), and a fiber-coupled homodyne-like interferometric detector (purple). Observed peak maxima corresponding to x, y, $\theta$, $\phi$, and z motion are denoted by vertical dashed lines. The x and y modes are degenerate for this B1 prism. The vertical scale corresponds to the displacement in the direction used for detector calibration, as indicated also by the bold label (see Supplementary Material). In some cases, higher harmonics of the labeled peaks are visible in the spectra, indicating the presence of non-linearities.}
\end{figure}

\textit{Observed dynamics-- }Power spectral density data for the motion of two sizes of hexagonal prisms, lower aspect ratio B1 and higher aspect ratio B2, trapped using $475$ mW of laser power at a vacuum of $2.2$ mbar are shown in Fig. \ref{fig:all_data} for each of the detection mechanisms described previously. 
We pinpoint the mechanical resonances of the prisms by identifying peaks at same frequencies in multiple detectors. Peaks corresponding to primary translational and rotational 
mechanical resonances $x$, $y$, $\theta$, $\phi$, and $z$ are denoted by vertical dashed lines. All other peaks are identified as 
known harmonics, sidebands, or electronic noise in the system. The conversion to nm/$\sqrt{\text{Hz}}$ is done for a harmonic trapping potential by assuming thermal equilibrium and fitting a strongly coupled peak to a Lorentzian function (see Supplementary Material). The thermally-driven torsional root-mean-squared amplitudes of the modes $\theta$ and $\phi$ can  be estimated by the equipartition theorem: $\theta(\phi) = \frac{1}{2\pi f_{\theta(\phi)}} \sqrt{\frac{k_{B}T}{I}}$, where $f_{\theta(\phi)}$ is the observed torsional frequency, $k_{B}$ is the Boltzmann constant, $T$ is the temperature, and $I$ is the moment of inertia of a B1/B2 hexagonal prism. The calculated root-mean-square amplitudes of $\theta$ and $\phi$ are 1.7$\times 10^{-2}$ rad and 1.5$\times 10^{-2}$ rad respectively for B1 and 9.1$\times 10^{-3}$ rad and 4.7$\times 10^{-3}$ rad for B2. 
We do not observe free rotations or bending modes, tested up to $10$ MHz, as expected since the $\phi$ and $\theta$ modes of the prism are small amplitude, stable linear oscillatory modes about the plane of the standing wave's interference fringes and bending modes are not predicted to be observable within our resolution for this geometry. Since no free rotations are observed, we also do not observe the presence of a significant precession term as in Ref. \cite{rashid2018precession}.

\textit{Mode Splitting--.}
The analytic model for a thin disc from Ref. \cite{seberson2020stability} predicts a small amount of mode splitting along x/y and $\theta/\phi$ axes, with the effect increasing the further the levitated object is from the Rayleigh regime - specifically in the radial axis. Therefore a wider hexagonal prism will have a less degenerate set of $\theta/\phi$ and x/y modes. This effect can be seen in both the outputs of the numerical model and the experimental data. Specifically in Fig. \ref{fig:all_data}, the x/y modes for the radially smaller B1 hexagon are nearly degenerate and the $\theta/\phi$ modes are much closer together than for the radially larger B2 hexagon.
Mode splitting is also often seen in levitated particles in the Rayleigh limit, caused by the asymmetric potential generated by polarisation dependent focusing of the beam  \cite{novotny2012principles}. In our case, since our levitated $\beta-$NaYF prisms are well into the Mie-Lorentz regime, separation between the modes results both from the polarization dependent focusing bias of the lens and geometry-dependent scattering resonances within the levitated particle \cite{petercrystalgrowth}. 


\textit{Finite element comparison-- }Fig. \ref{fig:optics_layout}d shows the PSD of an optically trapped hexagonal prism with values from the finite element model overlaid. The location of the peaks in the PSD is broadly in line with that computed by the finite element model. It is of note that the frequency of the $z$ DOF is mainly dependent on the thickness of the hexagonal prism, while the mode splitting is mainly dependent on its radius. 
We assume the prism has uniform thickness, 
and surface roughness is not modelled. 
Differences in the observed frequency of the $x/y$ motional modes to those predicted by the FEM model may be driven by edge effects due to the prism's tapered shape (see Fig. \ref{fig:sem_image} in Supplementary Material). Due to memory and CPU time driven resolution constraints the FEM model assumes a flat face on the hexagons prism's thin edge. Differences may also result due to geometry dependent Fabry-Perot and whispering gallery-mode-like resonances within the levitated prism \cite{petercrystalgrowth}.

\textit{Sensitivity to gravitational waves--}
For the application of gravitational wave detection, a disc-shaped particle or hexagonal prism can be suspended in a standing wave in an optical cavity. For such a configuration, as discussed in Ref. \cite{aggarwal2022searching}, the minimum detectable strain $h_{\rm{limit}}$ for a particle with center-of-mass temperature $T_\mathrm{CM}$ is approximately 
\begin{equation}
h_{\mathrm{limit}}=\frac{4}{\omega_0^2L}\sqrt{\frac{k_BT_{\mathrm{CM}}\gamma_gb}{M}\left[1+\frac{\gamma_{\mathrm{sc}}}{N_i\gamma_g}\right]}H\left(\omega_0\right), \label{eq:noise}
\end{equation}
for a particle trapped in a cavity with response function $H\left(\omega\right) \approx \sqrt{1+4\omega^2/\kappa^2}$ 
for a cavity of linewidth $\kappa$. Here $N_i=k_BT_\mathrm{CM}/\hbar\omega_0$ is the mean initial phonon occupation number of the center-of-mass motion.  $\gamma_g = \frac{32P}{\pi \bar{v} \rho t}$ is the gas damping rate at pressure $P$ with mean gas speed $\bar{v}$ for a plate or disc of thickness $t$ and density $\rho$, and $b$ is the bandwidth. 

\begin{figure}[!t]
\includegraphics[width=0.90\columnwidth]{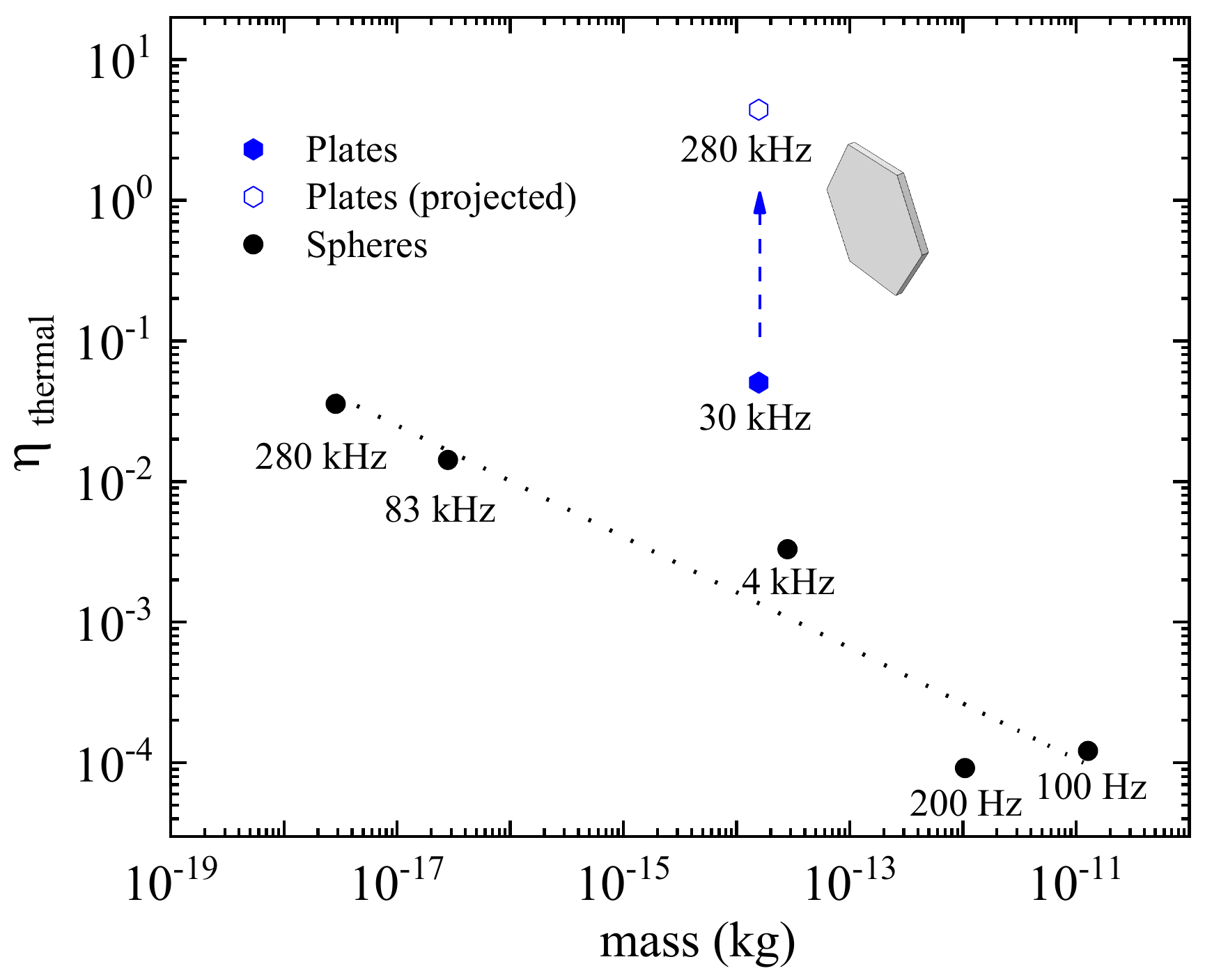}
\caption{\label{fig:mass_frequency_product} Comparison of the figure of merit $\eta_{\mathrm{thermal}} = \omega_0^2\sqrt{Ml}$ for sensitivity to gravitational waves in the thermal-noise dominated limit, for spheres and plates of mass $M$ and radius $r=l$ or thickness $t=l$, respectively, for recent experimentally realized trapping configurations. At equal masses, high aspect ratio levitated plates (blue hexagons, $5$ $\mu$m diameter prism from sample B2) significantly outperform levitated spheres (black circles, as reported in Refs. \cite{Uros}, \cite{smooshing}, \cite{atherton2015}, \cite{monteiro2017optical} , and \cite{monteiro2020search}, in order of decreasing frequency) for gravitational wave experiments due to their correspondingly higher trapping frequencies. Projected sensitivity for the same size plate held at higher trapping frequency is shown (open blue hexagon), as may be possible by using a higher intensity trap.}
\end{figure}

\begin{figure}[h]
\vspace{3mm}
\includegraphics[width=0.90\columnwidth]{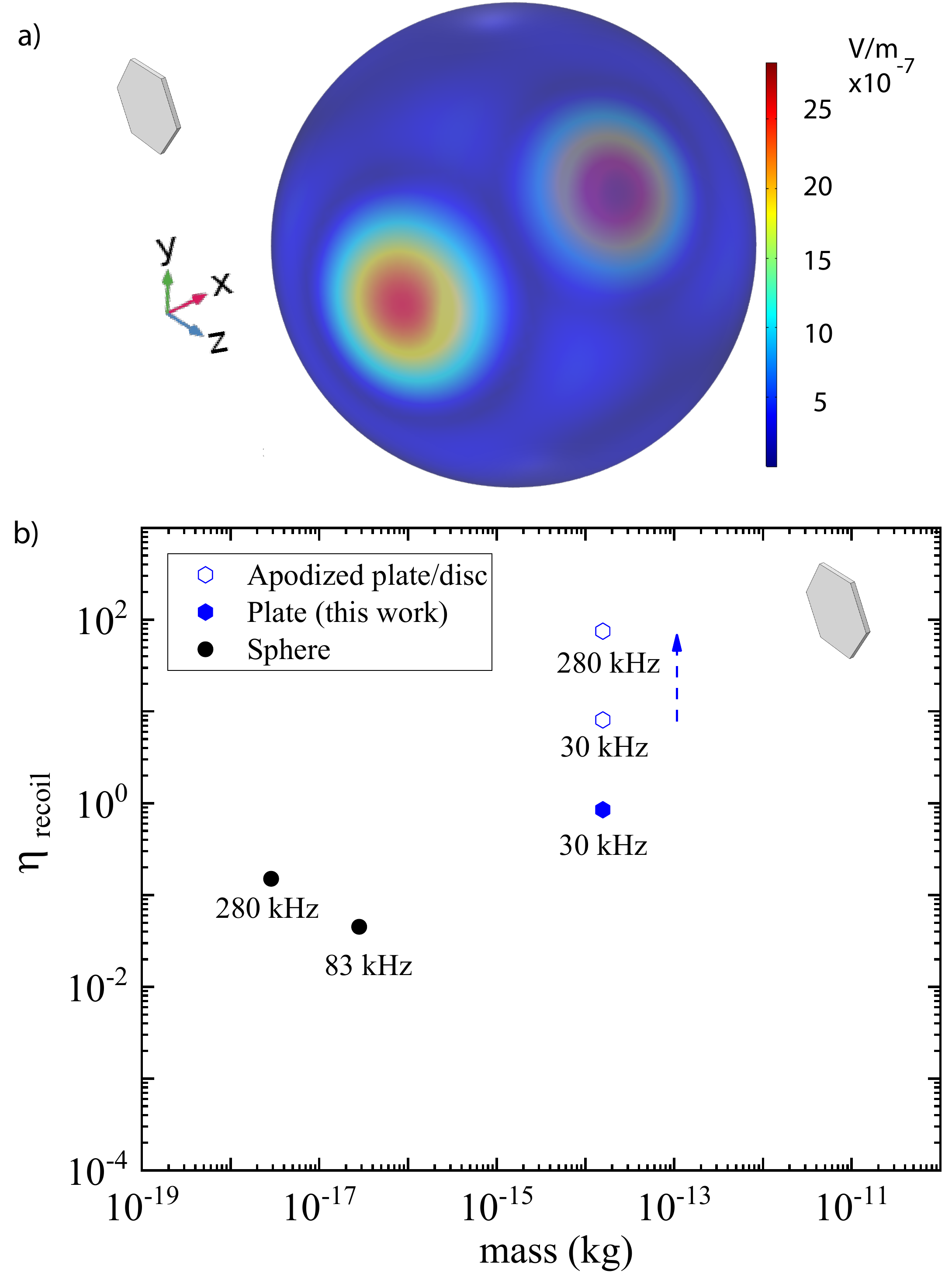}
\caption{\label{fig:recoil} (a) Calculated far-field scattered electric field profile $|E| \hat{k}$ for a hexagonal prism of thickess 200 nm and diameter $4$ $\mu$m (within the size range of the B2 samples), normalized to an incident electric field of $1$V/m. Incident beam propagates along $x-$ direction. Colorbar shows magnitude of field and surface plot shows directional dependence of scattering.  (b) Comparison of the figure of merit $\eta_{\mathrm{recoil}} = \omega_0^{3/2}\sqrt{M/\gamma_{sc}}$ for sensitivity to gravitational waves in the photon-recoil-heating dominated limit, for spheres and plates of mass $M$. 
Data for nanospheres is taken from recent experiments \cite{Uros,smooshing}. Projections shown for a prism of the size trapped in this work, for currently realized beam parameters with an estimated disc-limited finesse of $1.1 \times 10^3$ (solid hexagon) and for two trapping frequencies with a suitably apodized edge and beam waist profile, with an estimated disc limited finesse of $10^5$ (open hexagons).}
\end{figure}

The photon recoil heating rate for a plate- or disc-shaped structure is \cite{Arvanitaki:2013,Chang:2012,aggarwal2022searching}
$\gamma_{sc}=\frac{V_c\lambda\omega_0}{4L}\frac{1}{\int{dV(\epsilon-1)}}\frac{1}{\mathcal{F}_{\rm{disc}}}
$ is inversely proportional to the disc-limited finesse $\mathcal{F}_{{\rm{disc}}}$, i.e. \(2\pi\) divided by the fraction of photons scattered by the disc outside the cavity mode. The integral is performed over the extent of the suspended particle. Here $V_c$ is the cavity mode volume \cite{Arvanitaki:2013}.  For a sub-wavelength spherical particle of volume $V$ in the Rayleigh regime, $\gamma_{sc} = \frac{2}{5} \frac{\pi^2 \omega_0 V}{\lambda^3} \frac{(\epsilon-1)}{(\epsilon+2)}$.

Examining Eq. \ref{eq:noise}, we can identify two distinct regimes depending on the background pressure and trapping laser intensity. At higher pressure, the sensitivity tends to remain in the gas dominated regime, while at ultra-high vacuum, $\gamma_{sc}/N_i\gamma_g > 1$ and eventually the sensitivity is limited by photon recoil heating.
 We can define a figure of merit in the gas-dominated regime $\eta_{\mathrm{thermal}} = \omega_0^2\sqrt{Ml}$, where $\omega_0$ is the trapping frequency along the cavity axis, $M$ is the mass of the trapped object, $l=r$ or $l=t$ for spheres of radius $r$ or plates of thickness $t$, respectively.  Correspondingly, we can define figure of merit $\eta_{\mathrm{recoil}} = \omega_0^{3/2}\sqrt{M/\gamma_{sc}}$ for sensitivity to gravitational waves in the photon-recoil-heating dominated limit.
 
In Fig. \ref{fig:mass_frequency_product}, we show $\eta_{\mathrm{thermal}}$ for several recent experimentally realized configurations in various groups as well as for the particles trapped in this work. When comparing objects of similar mass, the higher frequencies obtained for our hexagonal prisms show a significant improvement.  Although the values of $\eta_{\mathrm{thermal}}$ realized in the present work are only marginally better than those obtained for some of the smallest nanospheres trapped 
at very high frequencies (driven by the $\omega_0^2$ scaling), by raising the trapping intensity we estimate for our geometry it should be possible to greatly exceed the highest values obtained in these experiments.  
 
 For operating in the ultra-high vacuum regime where we expect to be limited by photon recoil heating rather than gas damping, it is clear the disk or hexagonal prism shaped objects show a distinct advantage, due to the more-directional nature of their scattering (as shown in Fig. \ref{fig:recoil}a) in contrast to the nearly isotropic scattering from small spherical particles in the Rayleigh regime.  In Fig. \ref{fig:recoil}b we plot $\eta_{\mathrm{recoil}}$, focusing on the higher frequency domain with a comparison between nanoparticles in the Rayleigh limit and hexagonal prisms of thickness $200$ nm and diameter $5$ $\mu$m. 
 The hexagonal prisms greatly outperform the nanoparticles for the figure of merit $\eta_{\mathrm{recoil}}$ by two or more orders of magnitude for the case of a disc-limited finesse of $\mathcal{F}_{\mathrm{disc}}=10^5$, reasonable for a suitably apodized disc or prism trapped in a beam with waist equal to half the particle radius \cite{Chang:2012,aggarwal2022searching}.


\textit{Conclusion--}
In conclusion, we present the first optical trapping of very high aspect ratio disk-like Yb-doped $\beta-$NaYF hexagonal prisms, and compare the observed dynamics to a numerical model.  
The material we study in this work has also been shown in other experiments to exhibit anti-stokes fluorescence cooling, when illuminated with light of a suitable wavelength. Quasi-spherical nanocrystals with irregular morphologies prepared through top-down milling have also been cooled while being optically levitated \cite{Rahman2017}, and future work will examine 
the feasibility of simultaneous laser refrigeration and optical levitation of disc-like objects. This could allow higher trapping intensities and frequencies, yielding higher sensitivity to space-time strain from high frequency gravitational waves well above 100 kHz. 

For these prisms or other disc shaped sensors, the 
figure of merit $\eta$ for sensitivity to gravitational waves significantly improves upon other spherical levitated systems both in the gas damping limited regime and the photon recoil heating limited regimes.   
Transferring these prisms, or objects of a similar geometry, into a Michelson-type 1 meter cavity instrument represents the next step in the technical roadmap of the Levitated Sensor Detector (LSD) experiment \cite{aggarwal2022searching} -- this is work being undertaken at present. The successful characterisation/commissioning of these high aspect ratio hexagonal prisms therefore represents a significant step forward in developing $>$10 kHz gravitational wave astronomy.

\textit{Acknowledgements--.}
We thank Francis Robicheaux for insightful discussions on the equations of motion for the disks.  AG, GW, and NA are supported in part by NSF grants PHY-1806686 and PHY-1806671, the Heising-Simons Foundation, the John Templeton Foundation, and ONR Grant N00014-18-1-2370. AG and SL are supported by the W.M. Keck Foundation. VK is supported by a CIFAR Senior Fellowship and through Northwestern University through the D.I. Linzer Distinguished University Professorship. NA is also supported by the CIERA Postdoctoral Fellowship from the Center for Interdisciplinary Exploration and Research in Astrophysics at Northwestern University. This work used KyRIC (Kentucky Research Informatics Cloud) through the Extreme Science and Engineering Discovery Environment (XSEDE) with allocation TG-PHY210012, and the Quest computing facility at Northwestern.

\nocite{*}

\bibliography{apssamp}

\section*{Supplementary Material}

\newcommand{\hbAppendixPrefix}{S}
\renewcommand{\thefigure}{\hbAppendixPrefix\arabic{figure}}
\setcounter{figure}{0}

\subsection{\label{sec:appendix_hex_fab} Hexagon growth and characterization}

$\beta$-NaYF crystals were hydrothermally grown in an autoclave (Parr Instrument Company). YCl3 (99.9$\%$), YbCl3 (99.998$\%$), ErCl3 (99.9$\%$), TmCl3 (99.99$\%$), and ethylenediaminetetraacetic acid (EDTA, $>$99$\%$) were purchased from Sigma-Aldrich. NaOH was purchased from Fisher Scientific.  NaF (99.5$\%$) was purchased from EMD Chemicals. A 90$\%$ YCl3 and 10$\%$ YbCl3 solution by molar ratio was prepared. ErCl3 and TmCl3 can be added to this solution as dopants for upconversion experiments. EDTA was dissolved in water with 3 equivalents of NaOH to make a 0.2 M tri-sodium EDTA solution. NaF was dissolved in deionized water to make a 1 M stock solution.  

To synthesize the $\beta$-NaYF, 5 mL of 0.2 M YCl3 (10$\%$ YbCl3) was dispensed into an autoclave liner and chelated with 5 mL of 0.2 M tri-sodium EDTA by stirring for 10 minutes. After this, 4 mL of 1 M NaF was added to the mixture and stirred vigorously for 30 minutes. The mixture was then transferred to an autoclave and heated in an oven at 220°C for 24 hours. The sample was collected and isolated by centrifugation and washed twice with water and twice with ethanol. The sample was then dried at 80°C for 12 hours.  

Fig. S1 shows a typical distribution of both single and clustered particles, as viewed with a scanning electron microscope. After the particles are shaken loose from a substrate fixed above the trapping region, both individual and clustered particles have been successfully loaded into the optical trap. The characteristic displacement power spectral densities (PSDs) for single hexagonal prisms are determined by both statistics (by trapping many and selecting recurring non unique PSD patterns) and matching the observed PSDs to kinematic theory. In general, optically trapped clusters outnumbered trapped singles at around a $\sim 2:1$ ratio. For our loading method we find this ratio decreases roughly in direct proportion to the age of the synthesised batch, over the timescale of a few months.

\begin{figure}[h]
\includegraphics[width=0.98\columnwidth]{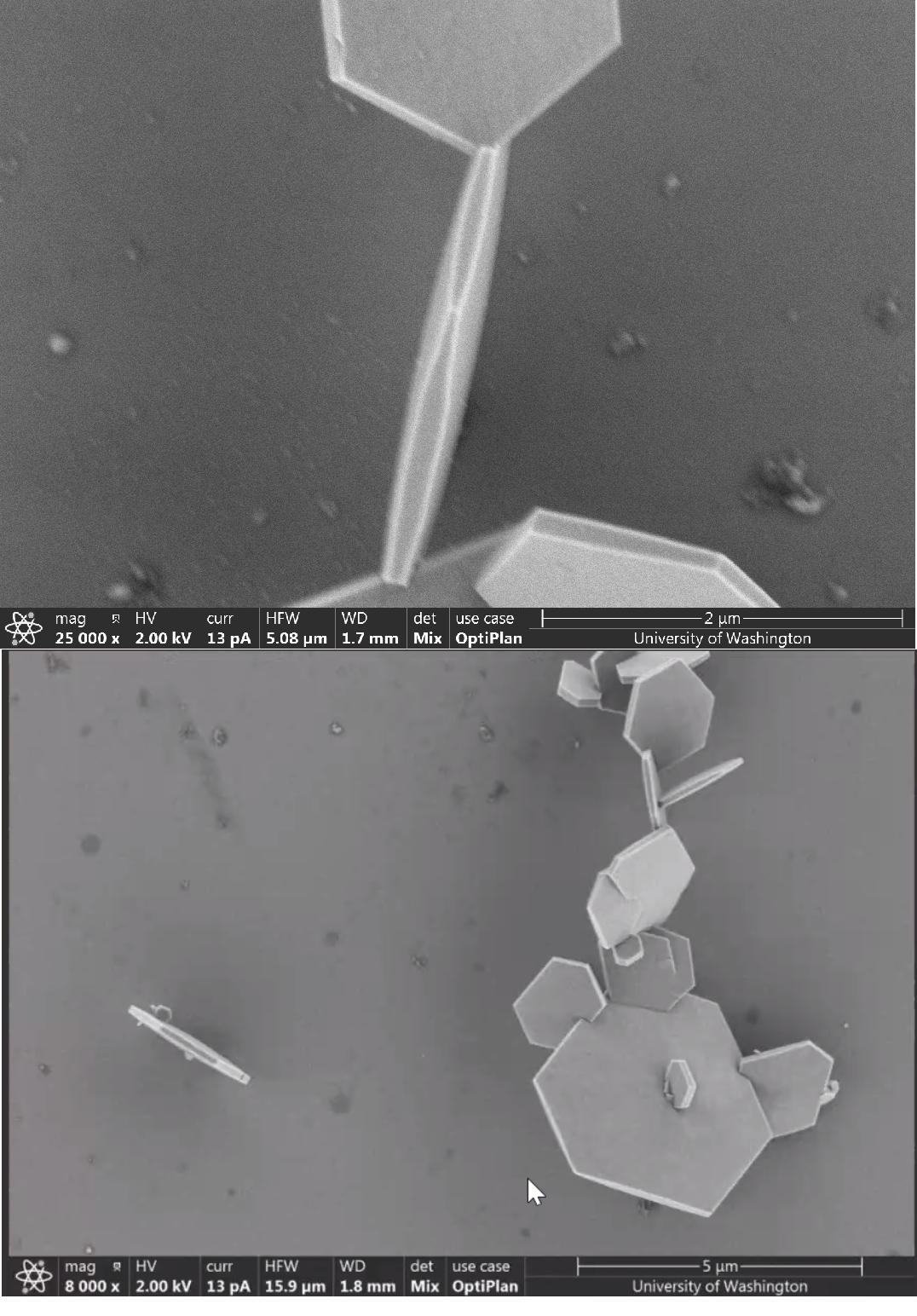}
\caption{\label{fig:sem_image} Scanning electron microscope images of NaYF hexagonal prisms. Note the presence of both 'single' particles  - left of bottom image - and 'clustered' particles - right of bottom image. Top image depicts slight curvature of the face of a prism as visible when viewed upon its edge.}
\end{figure}

\subsection{\label{sec:Detection_calibration} Detector displacement calibration}

For the data in Fig. \ref{fig:all_data} in the main text, the y-axis was scaled from arbitrary units of the detected optical signal to distances in nm by treating the particle as a harmonic oscillator subjected to a linear restoring force in thermal equilibrium. For each data set, one peak was chosen that corresponds to a translational motion that is well-coupled to that detector. The chosen peak at resonant frequency $\omega_0$ was fit to the form of the square root of the power spectral density (PSD) function for a harmonic oscillator $S_q(\omega)$ in terms of natural frequencies $\omega$,
\begin{equation}
S_q(\omega) = \frac{k_B T_b \gamma} {2 \pi^3 m[(\omega^2 - \omega_0^2)^2 + \omega^2\gamma^2]}, \label{eq:PSD}
\end{equation}
where $T_b$ is the bath temperature, $\gamma$ is the measured damping rate, and $m$ is the estimated mass of the particle. A scaling constant $C_{\mathrm{scaling}}$ was determined that relates the expected position-based PSD $S_q(\omega)$ to the PSD in volts $S_V(\omega)$ from the fit according to $S_q(\omega) = C_{\mathrm{scaling}}^2 S_V(\omega)$.

For B1, the $x/y$ degenerate peak was used for calibration of all side (i) and bottom (ii) detectors, while the $z$ peak was used for the diagonal (iii) and axial fiber-based (iv) detectors. For B2, the $x$ peak was used for the side (i) detectors, the $y$ peak was used for the bottom (ii) and diagonal (iii) detectors, and the $z$ peak was used for the axial fiber-based (iv) detector.

Since only one peak is used for calibration of each data set, amplitudes of additional translational peaks in each spectra are to be interpreted as the component of that motion projected in the optimal direction of the detector. 
Although the spectra are calibrated using motion that is strongly coupled to the sensitive direction of the corresponding detectors, in general the normal modes of oscillation in the trap are not necessarily perfectly aligned with the lab reference frame, particularly for the $x-$ and $y-$ translational degrees of freedom and torsional modes, and this contributes some uncertainty towards the displacement calibration for the corresponding dominant peaks. In general we expect $S_V = \sum_q \chi_q S_q$ for a set of scaling factors $\chi_q$.  In our approximation we consider only one value of $q$ (corresponding to the dominant peak) for each spectrum's voltage-to-displacement calibration, which could result in uncertainties of order $\sim 10-20$ percent.  More accurate calibrations could in principle be achieved by using a known electric field and charge on the trapped object (see e.g. Ref. \cite{ranjit2016zeptonewton} in the main text).

\begin{figure}[htp]
\includegraphics[width=0.99\columnwidth]{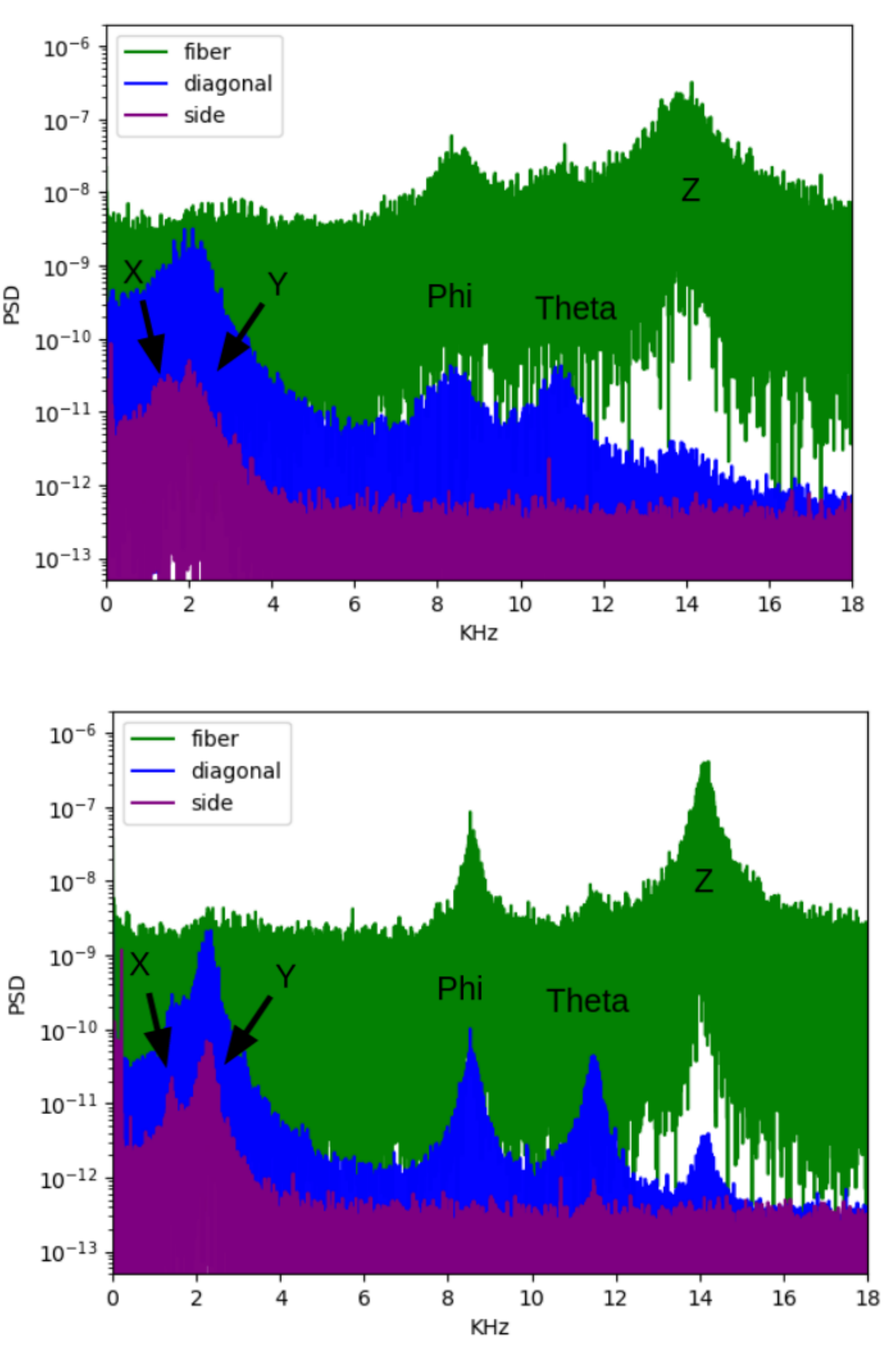}%
\caption{\label{fig:pressure} Experimentally obtained power spectral densities for two different pressures (3 and 12 mbar) of an optically trapped B1 hexagonal prisms from multiple detectors oriented to the side (purple), top diagonal (blue), and along the standing wave (trap) axis (green). The side (purple) and diagonal (blue) detectors are free space coupled biased photodiodes, the standing wave axis (green) is a fiber coupled balanced photodetector in a homodyne-like interferometric configuration. Observed peak maxima corresponding to x, y, $\theta$, $\phi$, and z motion are denoted by arrows. The x and y modes are (mostly) degenerate for this B1. 
}

\end{figure}

%
%
%
%
%


\subsection{Evolution with pressure }

Figure S2 shows the evolution of the trapped hexagon's PSDs with pressure. The cartesian/translation and vibrational degrees of freedom obtain higher $Q$ factor at lower pressures as expected. Free rotations or precessional degrees of freedom would change their central frequencies with pressure, however we do not observe these.


\end{document}